\documentclass[a4paper]{jpconf}
\usepackage{graphicx}
\usepackage{color}
\usepackage{bm}
\newcommand{\mib}[1]{\bm{#1}}
\begin{document}

\title{Valence Imbalance of Manganese Ions between Surface and Bulk Enhanced by Fermi-Surface Structure in Layered Manganites}

\author{Ryosuke Yamamura and Takashi Hotta}

\address{Department of Physics, Tokyo Metropolitan University, Hachioji, Tokyo 192-0397, Japan}

\ead{yamamura-ryosuke@ed.tmu.ac.jp}

\begin{abstract}
To investigate valence imbalance phenomena between the surface and bulk in layered manganites,
we analyze an $e_{\rm g}$-orbital degenerate double-exchange model with surfaces
for two types of $t_{\rm 2g}$ spin structures.
We reconfirm that the surface-induced Friedel oscillations occur in the charge structure
and that the number of Mn$^{4+}$ ions on the surface is larger than that in the bulk.
This tendency is found to be more significant,
when the $e_{\rm g}$-electron system has Fermi-surface structures with better nesting properties.
The behavior of the Friedel oscillations depends on the nesting properties of the Fermi-surface
curves along the direction of the oscillations, and not on the dimension of the system itself.
We believe that these results will be useful for the development of
high-efficiency cathodes in Li-ion batteries and catalysts.
\end{abstract}

\section{Introduction}

The development of rechargeable batteries has been an important milestone
in the quest for clean sources of energy.
Among batteries, Li-ion batteries with LiCoO$_{2}$ as the cathode material 
have significant high energy density and found remarkable success 
in application to portable electronic devices \cite{LiCoO2a,LiCoO2b}.
However, since LiCoO$_{2}$ suffers from high cost, toxicity, and safety issues,
the material is not appropriate for use in electric vehicles.
In recent decades, substitutional materials for LiCoO$_{2}$ have been
investigated, and some manganites are preferred for Li-ion batteries
since they have low cost, no toxicity, and better safety characteristics.
In fact, LiMn$_{2}$O$_{4}$ with a spinel structure has been successfully
employed as the cathode in electric vehicles \cite{LiMn2O4}.
LiMnO$_{2}$ and related materials including other transition metal elements
with a layered structure are promising cathode materials \cite{LiMnO2, LiNiMnO2}.
Manganites such as CaMn$_{4}$O$_{5}$ and MnO$_{2}$
are also expected to act as catalysts in photosystems and artificial photosynthesis
\cite{CaMn4O5,MnO2a,MnO2b}.

For application to cathodes and catalysts,
{\it surface} electron states play important roles
because manganites work as cathodes and catalysts in chemical reactions
involving a change of the valence of surface manganese ions mainly from Mn$^{4+}$ to Mn$^{3+}$.
Furthermore, in the case of cathodes, if Mn$^{3+}$ ions are present on the surface, 
manganese dissolution from the lattice into the electrolyte occurs owing to
the disproportion of Mn$^{3+}$ ions into Mn$^{2+}$ and Mn$^{4+}$ ions \cite{Choi}.
This reaction is one of the most important factors causing the capacity fading of cathodes in Li-ion batteries. 
Thus, for the material to achieve high efficiency as a cathode or catalyst,
it is desirable to retain the Mn$^{3+}$ ions in the bulk while the Mn$^{4+}$ ions reside on the surface.
Previous studies approached this problem using first-principles calculations
of the charge structure on the surface of manganites.
For $\lambda$-MnO$_{2}$, which is obtained by removing Li ions from LiMn$_{2}$O$_{4}$, 
a portion of Mn$^{4+}$ ions transition to Mn$^{3+}$ ions because of surface reconstruction \cite{Ouyang}.
First-principles calculations have enabled us to consider surface reconstruction or so,
but such calculations are limited to small clusters.
Therefore, studies of the microscopic mechanisms causing the valence imbalance of manganese ions 
between the surface and bulk are meaningful when trying to understand increasingly large clusters.

To further the above research,
we have investigated the effect of the surface on the electronic state
in layered manganites using a $1000 \times 1000 \times 100$ lattice 
on the basis of an orbital-degenerate double-exchange model with surfaces \cite{Yamamura}.
%
%
%
In Ref.~\cite{Yamamura},
in order to clarify the difference in the electron numbers between the surface and bulk,
we have concentrated on $e_{\rm g}$ electron states
for some fixed $t_{\rm 2g}$ spin patterns.
As typical examples, we have chosen two types of $t_{\rm 2g}$ spin structures: 
the C-type antiferromagnetic (AF) and sheet-like AF phases.
These phases were chosen because they have been found in the bulk 
for some values of $J$, when we include the effect of Jahn-Teller distortions
in the $2 \times 2 \times 2$ and $4 \times 4 \times 4$ lattices \cite{Hotta1,Hotta2}.
In those cases, we have confirmed that the number of Mn$^{4+}$ ions on the surface is
always larger than that in the bulk,
which is understood by surface-induced Friedel oscillations.
Note that the behavior of Friedel oscillations is different between two $t_{\rm 2g}$
spin structures.
Thus, it is necessary to carefully investigate the behavior of these oscillations
for future applications to cathodes and catalysts.

In this paper, we investigate the behavior of surface-induced Friedel oscillations
in layered manganites using an orbital-degenerate double-exchange model
with surface effects for other types of $t_{\rm 2g}$ spin structures.
It is found that the number of Mn$^{4+}$ ions on the surface is always
larger than that in the bulk owing to surface-induced Friedel oscillations.
Furthermore, irrespective of the type of $t_{\rm 2g}$ spin structures,
we find that the Friedel oscillations are enhanced
when the Fermi-surface curves with better nesting properties exist.
Namely, the dimension of the area of ferromagnetic (FM) $t_{\rm 2g}$ spin structure
is not essential to the enhancement of Friedel oscillations.
We believe that these results will contribute to the development of
highly-efficient cathodes and catalysts.
For the calculations in this paper, we use $k_{\rm B}=\hbar=1$.

\section{Model and Method}

Herein we introduce the orbital-degenerate double-exchange model \cite{Zener}.
This model has been widely used to successfully investigate the physical properties 
in layered manganites, such as colossal magneto-resistance phenomena \cite{Tokura} 
and various types of charge and/or orbital orders in the bulk \cite{Hotta1,Hotta2,Dagotto1,Dagotto2}.
To simplify the model, we consider a widely-used approximation
that invokes infinite Hund coupling between $e_{\rm g}$ and
$t_{\rm 2g}$ electron spins.
In this limit, the $e_{\rm g}$ electron spin perfectly aligns along
the $t_{\rm 2g}$ spin direction,
reducing the spin degrees of freedom of $e_{\rm g}$ electrons.
Namely, the $e_{\rm g}$ electrons move only in the FM region of $t_{\rm 2g}$ spins,
while in the AF region, $e_{\rm g}$ electrons are localized.
The model Hamiltonian for this system is given by
\begin{equation}
 H =  -\sum_{\mib{i} \mib{a} \gamma \gamma'}
        D_{\mib{i},\mib{i}+\mib{a}} t^{\mib{a}}_{\gamma \gamma'}
        d^\dagger_{\mib{i}\gamma}d_{\mib{i}+\mib{a}\gamma'} 
     + J \sum_{\langle \mib{i},\mib{j} \rangle} S_{z \mib{i}} S_{z \mib{j}},
\end{equation}
where $d_{\mib{i} a}$ ($d_{\mib{i} b}$) is the annihilation operator
for a spinless $e_{\rm g}$ electron in the $d_{x^{2}-y^{2}}$
($d_{3z^{2}-r^{2}}$) orbitals at site $\mib{i}$,
$\mib{a}$ is the vector connecting nearest-neighbor sites,
$D_{\mib{i},\mib{i}+\mib{a}}=(1+S_{z \mib{i}}S_{z \mib{i}+\mib{a}})/2$,
$S_{z\mib{i}}$ is an Ising-like $t_{\rm 2g}$ spin at site $\mib{i}$
with $S_{z\mib{i}}=\pm 1$,
and $t^{\mib{a}}_{ \gamma \gamma'}$ denotes the nearest-neighbor
hopping amplitude between $\gamma$ and $\gamma'$ orbitals
along the $\mib{a}$ direction with
$t^{\mib{x}}_{aa}=-\sqrt{3}t^{\mib{x}}_{ab}
=-\sqrt{3}t^{\mib{x}}_{ba}=3t^{\mib{x}}_{bb}=3t^{\mib{z}}_{bb}/4$,
$t^{\mib{y}}_{aa}=\sqrt{3}t^{\mib{y}}_{ab}
=\sqrt{3}t^{\mib{y}}_{ba}=3t^{\mib{y}}_{bb}=3t^{\mib{z}}_{bb}/4$,
and $t^{\mib{z}}_{aa}=t^{\mib{z}}_{ab}=t^{\mib{z}}_{ba}=0$ 
%
\cite{Slater}.
We set the energy, $t^{\mib{z}}_{bb}=1$.
We note that the hopping amplitudes between different orbitals 
of a spinless $e_{\rm g}$-electron system
change sign between along the x-and y- directions,
$t^{\mib{x}}_{ab}=t^{\mib{x}}_{ba}=-t^{\mib{y}}_{ab}=-t^{\mib{y}}_{ba}$.
The second term in Eq. $(1)$, $J$, is the AF coupling between
the nearest-neighbor $t_{\rm 2g}$ spins and
$\langle \mib{i},\mib{j} \rangle$ denotes the nearest-neighbor site pair.
In this paper, we do not include explicitly the Coulomb interaction terms,
but the effect is partially considered by working with a large Hund coupling.
Namely, we consider a manganite with large bandwidth.
We define $x$ as the hole doping
and $n$ as the electron number of $e_{\rm g}$ electrons per site 
with a relation of $x=1-n$.
Thus, $x$ indicates the average number of Mn$^{4+}$ ions per site.

We diagonalize the Hamiltonian
in a $1000 \times 1000 \times 100$ lattice:
one layer is composed of a $1000 \times 1000$ square lattice
with periodic boundary conditions along the x- and y-axes.
To introduce surfaces to the model, we consider a stack of
$100$ layers with open boundary condition (OBC) along the z-axis.
In this paper, we introduce the OBC in the sense of $\psi (0)=\psi (101)=0$,
where $\psi(z)$ denotes the wavefunction of a spinless $e_{\rm g}$ electron
along the z-axis and $z$ is a layer number.
Namely, the first layer along the z-axis is the surface,
whereas the $50$-th layer denotes the bulk.

\begin{figure}[t]
\centering
\includegraphics[width=\columnwidth,keepaspectratio]{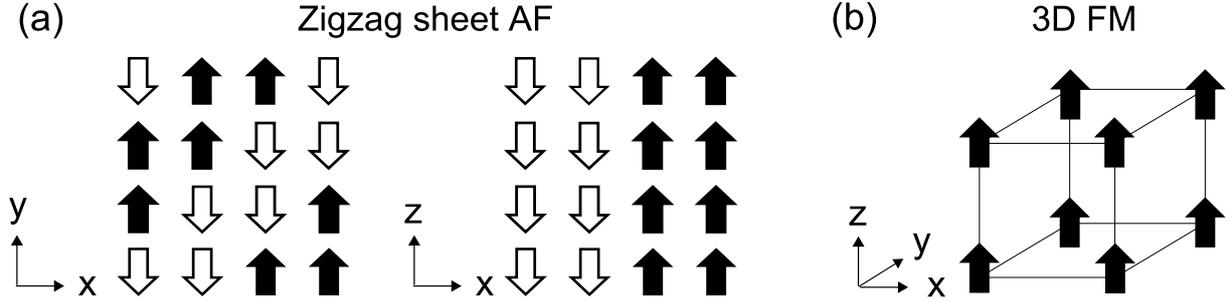}
\caption{
Schematic views of spin structures cited in this paper.
Solid and open arrows indicate up and down $t_{\rm 2g}$ spins, respectively.
(a) The spin structure of the zigzag sheet AF state.
(b) The spin structure in the FM state.
}
\label{fig:fig1TMU}
\end{figure}

In this paper,  we remark the $e_{\rm g}$-electron properties
on the background of the fixed $t_{\rm 2g}$ spin structure.
In our previous study, we investigated two $t_{\rm 2g}$ spin structures:
a C-type AF (C-AF) state and an A-type AF (A-AF) state \cite{Yamamura}.
In the C-AF state, $e_{\rm g}$ electrons can move only
along the z-axis owing to infinite Hund coupling.
Thus, the system is regarded as a one-dimensional (1D) chain with the OBC along the z-axis.
In the A-AF state, $e_{\rm g}$ electrons can move in the x-z plane,
leading to a two-dimensional (2D) sheet with edges.
On the other hand, in this paper, 
we assume two other possible $t_{\rm 2g}$ spin structures,
zigzag sheet AF and FM states, as shown in Figs.~1(a) and 1(b), respectively.
At least in a $\sqrt{8} \times \sqrt{8} \times 3$ lattice 
%
with surfaces, 
it is easy to show that
these magnetic states actually appear as the ground states for appropriate values of $J$,
but detailed discussion on the stability of ground-state magnetic structures
in the large cluster is postponed in future. 
In the zigzag sheet AF state, 
the FM regions of $t_{\rm 2g}$ spins form zigzag structures
in the x- and y- directions, 
whereas $t_{\rm 2g}$ spins are ferromagnetic in the z- direction.
Thus, the system is regarded as a 2D sheet with edges,
where hopping amplitudes change periodically.
In the three-dimensional (3D) FM state,
$e_{\rm g}$ electrons can move in all three dimensions.

By using these lattices, we evaluate $N(z)$,
the number of Mn$^{4+}$ ions per site in layer $z$, and
investigate the behavior of the surface-induced Friedel oscillations.
For several $t_{\rm 2g}$ spin structures with different FM region dimensions, 
we examine the factor that determines the behavior of the oscillations.
Note that $N(1)$ and $N(50)$ denote the number of 
Mn$^{4+}$ ions per site on the surface and in the bulk, respectively.

\section{Calculation Results}

\subsection{Brief review of previous results}

Before proceeding to the results of our calculations,
we need to briefly review our previous results \cite{Yamamura}.
We observe that $N(z)$ clearly exhibits a damped oscillation from the surface ($z=1$)
to the bulk ($z=50$), which is understood to be the surface-induced Friedel oscillation,
and results in an increase in the number of Mn$^{4+}$ ions on the surface.
In the 1D system such as the C-AF state, the formula for Friedel oscillation is expressed
as \cite{Friedel,Shibata,FO1,FO2,FO3}
\begin{equation}
 \label{eq:1D}
  N(z)-x \propto \cos({2k_{\rm F}z}) z^{-(1+K_{\rho})/2},
\end{equation}
where $k_{\rm F}$ is the Fermi momentum
and $K_{\rho}$ is the correlation exponent \cite{Schulz}.
Since we do not include Coulomb interactions among $e_{\rm g}$ electrons,
$K_{\rho}=1$, which leads to $z^{-1}$.

On the other hand, in the 2D system such as the A-AF state,
the formula for Friedel oscillations is modified to
\begin{equation}
  \label{eq:2D}
  N(z)-x \propto \cos({2k_{\rm Fz}z}) z^{-C(x,\gamma)},
\end{equation}
where $k_{\rm Fz}$ is the z-component of the Fermi momentum
with good nesting properties along the z-axis.
$C(x, \gamma)$ is the damping power for the oscillations.
We have confirmed that in the 2D system
$C$ depends on hole doping, $x$, and a relevant orbital, $\gamma$, 
which leads to $C(x, \gamma) \geq 1$.
In other words, the envelope of the 2D Friedel oscillations decays faster than $z^{-1}$.
At this stage, the definite expression of $C(x, \gamma)$ is not obtained.
However, in general, $N(1)$ seems to be increasingly suppressed in higher dimensional systems
since the influence of the surface on the system becomes relatively weaker 
as the dimension of the FM region increases.

\begin{figure}[t]
\centering
\includegraphics[width=\columnwidth,keepaspectratio]{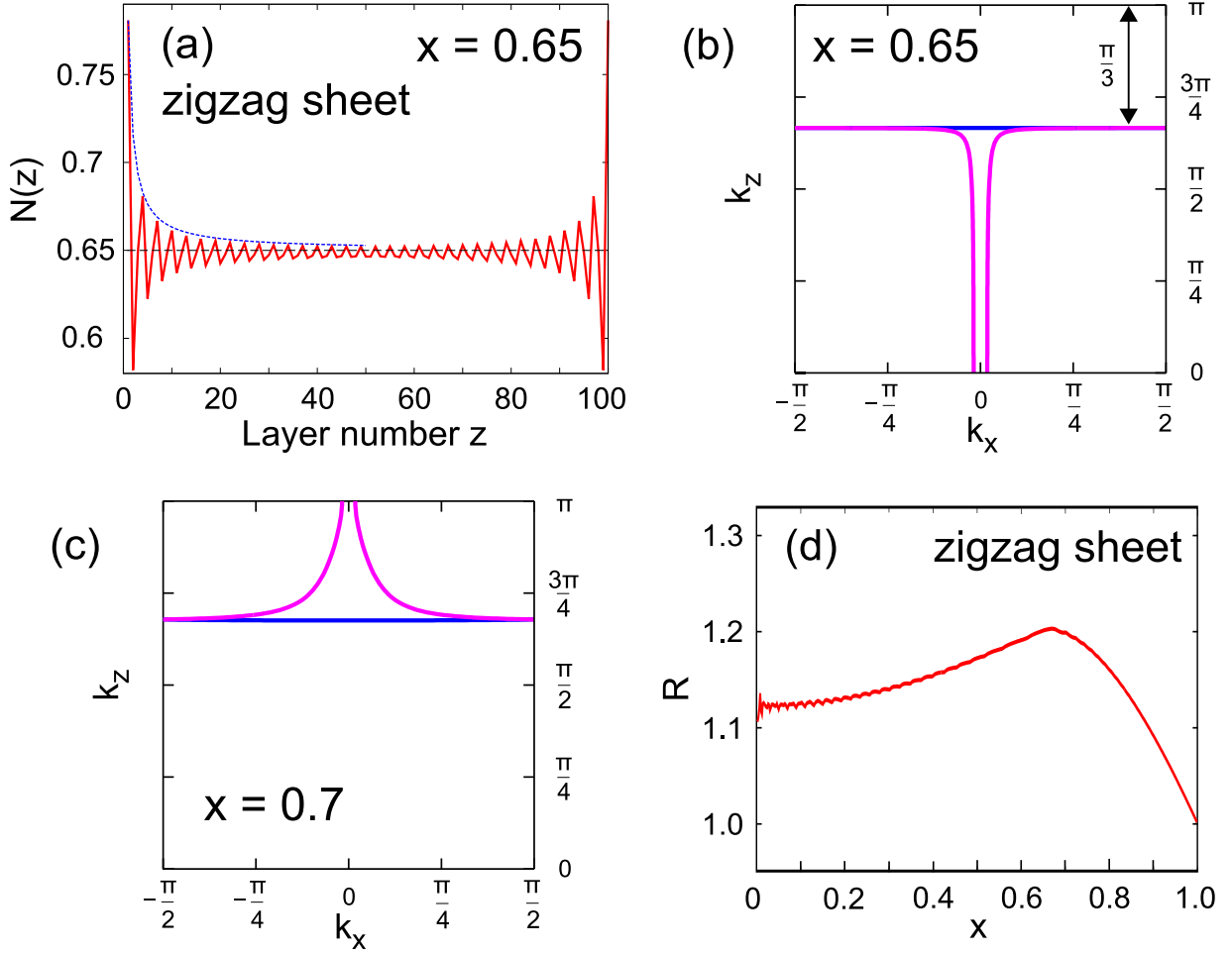}
\caption{
(a) $N(z)$ vs. $z$ plot for $x=0.65$ in the zigzag sheet AF state.
the blue dotted curve denotes the envelope expressed by $z^{-1}$,
and the broken horizontal line denotes the value of $x$.
%
(b) Two Fermi-surface curves of the $e_{\rm g}$-electron system for $x=0.65$.
The magnitudes of the Fermi momentum along the z-axis are both $\pi/3$,
which is consistent with the period of the oscillation in (a).
(c) Two Fermi-surface curves for $x=0.7$.
(d) Efficiency $R$ vs. $x$. $R$ is defined as $R=N(1)/x$.
}
\label{fig:fig2TMU}
\end{figure}

\subsection{Zigzag sheet AF state}

First, let us consider the charge structure in the zigzag sheet AF state.
In Fig.~2(a), we show $N(z)$ vs. $z$ for $x=0.65$.
Since the zigzag sheet AF indicates a 2D-like system 
even though the FM region is not a plane (see Fig.~1(a)), 
one could naively consider $N(z)$ to be expressed by Eq.~(\ref{eq:2D}).
However, as observed in Fig.~2(a), the envelope of the oscillations is well
expressed by $z^{-1}$.
Namely, the Friedel oscillation pattern is rather 1D-like in spite of the 2D-like FM region.

To investigate one of the factors that determines the value of $C(x, \gamma)$,
let us examine the oscillations from a viewpoint of the Fermi-surface structure.
In Fig.~2(b), we show  
%
two 
Fermi-surface curves at $x=0.65$.
Note that we depict the Fermi-surface curves considering OBC along the z-axis.
When we consider the $100$ layers stacked along the z-axis with OBC,
$k_z$ is given by $k_z(\ell)=\ell\pi/101$ where $\ell$
($=1, 2, \cdots, 100$) is an integer, leading to $\delta k_z=\pi/101$.
In the zigzag sheet AF state, we find that the spinless $e_{\rm g}$-electron system
possesses a quasi-1D Fermi-surface structure along the z-axis for a certain hole doping, $x$.
Even in the 2D system, when the Fermi-surface curves have a quasi-1D structure
along the z-axis,  the envelope of the oscillations is expressed by $z^{-1}$ as if in the 1D system.
The period of the oscillations in Fig.~2(a) is given by  $2 \times \pi/3$, 
consistent with the $k_{\rm Fz}$ shown in Fig.~2(b) and used in Eq.~(\ref{eq:2D}).
Fig.~2(c) shows the Fermi-surface curves at $x=0.7$.
For $x=0.7$, the Fermi-surface curves also retain the quasi 1D structure along the z-axis 
and we can confirm the value of $C(x, \gamma)$ is almost unity.
The nesting properties are lost for $x < 0.65$ and $x > 0.7$, 
so the envelope of the oscillations decays faster than $z^{-1}$.

To understand the increase in $N(1)$ compared with $x$,
we evaluate the efficiency $R$, given by $R=N(1)/x$ \cite{Yamamura}.
This quantity denotes to what degree the
Mn$^{4+}$ ions appear on the surface at a given hole doping $x$.
When $N(1)$ is enhanced, we expect that the efficiency
as cathodes and catalysts is increased.
Thus, $R$ has been called the ``efficiency''.
Note that $R$ is associated with the magnitude of amplitude of the Friedel oscillations.
Fig.~2(d) shows the plot of $R$ as a function of $x$
for the zigzag sheet AF state.
We find the magnitude of $R$ exhibits a peak between $x = 0.65$ and $0.7$,
%
for which one Fermi-surface curve has perfect nesting properties,
leading to a Van Hove singularity.
Note that the small oscillations in $R$ around $x=0.0$ are effects 
due to the size of the lattice.
If we increase the size of the lattice, these oscillations disappear.

Thus, we can conclude that 
when the spinless $e_{\rm g}$-electron system possesses
a Fermi-surface structure with decent nesting properties along the z-axis,
the value of $C(x, \gamma)$ becomes unity
and the amplitude of the Friedel oscillations tends to be larger.
Namely, the behavior of the Friedel oscillations is $not$ determined
by the dimension of $t_{\rm 2g}$ spin structure itself,
but the Fermi-surface structure of $e_{\rm g}$-electron system
is essentially important.


\begin{figure}[t]
\centering
\includegraphics[width=\columnwidth,keepaspectratio]{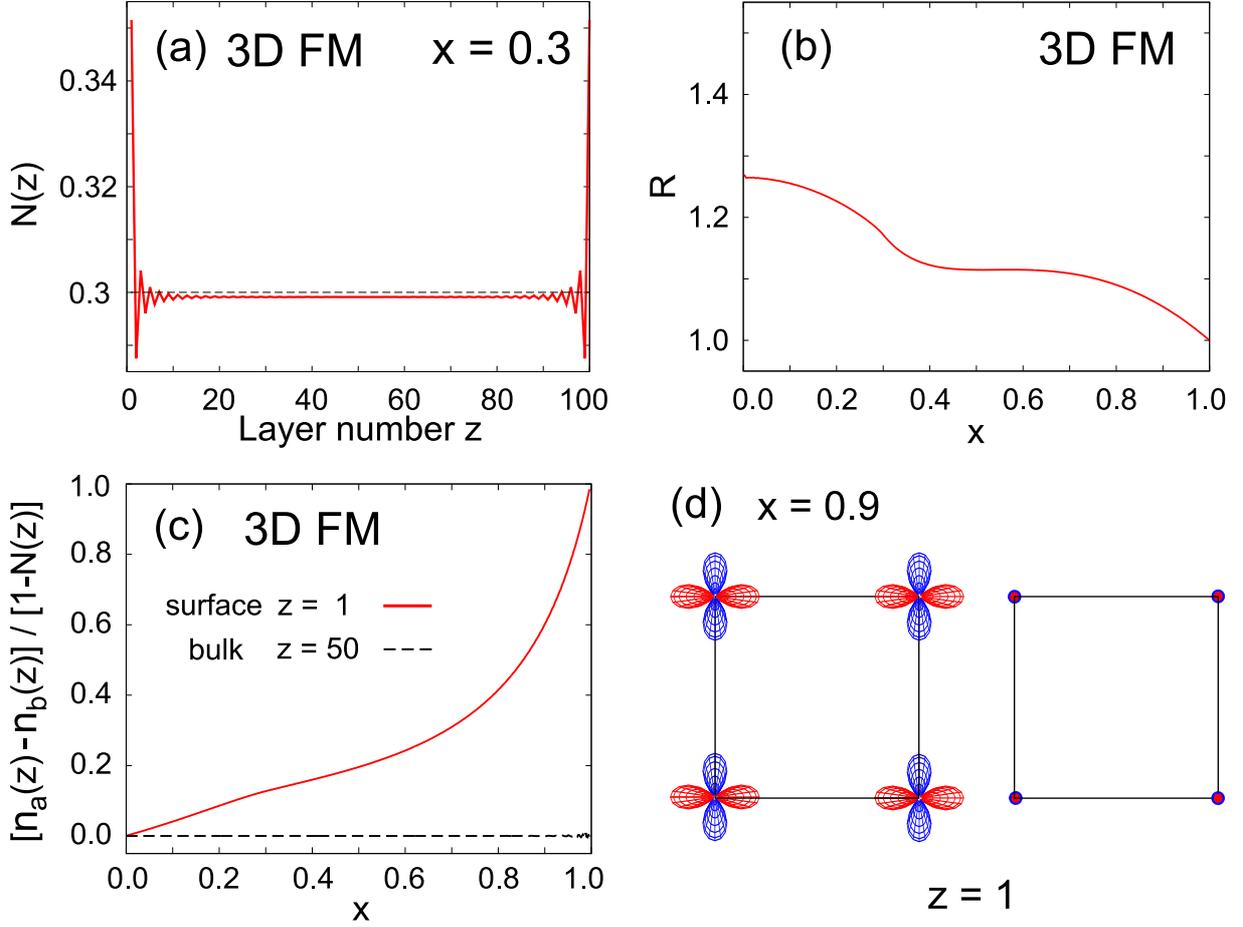}
\caption{
(a) $N(z)$ vs. $z$ plot for $x=0.3$ in the 3D FM state.
the broken horizontal line denotes the value of $x$.
(b) Efficiency $R$ vs. $x$.
(c) $[n_a(z)-n_b(z)]/[1-N(z)]$ vs. $x$ for $z=1$ and $50$.
(d) Orbital arrangement on the surface x-y plane for $x=0.9$.
The orbital size is chosen to be proportional to the orbital density.
}
\label{fig:fig3TMU}
\end{figure}

\subsection{3D FM state}

Fig.~3(a) shows the plot of $N(z)$ vs. $z$ for $x=0.3$;
surface-induced Friedel oscillations can be observed in the figure.
Note that $N(50)$ is slightly smaller than $x$.
In contrast to the case of the zigzag sheet AF state, 
the envelope of the oscillations decays faster than $z^{-1}$, 
since the nesting properties are weak in the 3D $e_{\rm g}$-electron system.
As observed in Fig.~3(b), $R$ decreases monotonically in the FM state, 
which can be explained by the fact that the nesting properties in the 3D $e_{\rm g}$-electron system 
are enhanced to some extent for smaller $x$.

A characteristic point for the FM phase is the appearance of
an orbital arrangement on the surface,
in sharp contrast to other $t_{\rm 2g}$ spin structures.
In Fig.~3(c), we plot $[n_a(z)-n_b(z)]/[1-N(z)]$ vs. $x$
for $z=1$ and $50$,
where $n_a(z)$ and $n_b(z)$ denote the average electron
number in the $d_{x^{2}-y^{2}}$ and $d_{3z^{2}-r^{2}}$ orbitals,
respectively.
In the bulk, $d_{x^{2}-y^{2}}$ and $d_{3z^{2}-r^{2}}$
orbitals are occupied with equal weights because
cubic symmetry is recovered in the 3D environment.
On the surface, $d_{x^{2}-y^{2}}$ orbitals are selectively occupied,
since $e_{\rm g}$ electrons on the surface layer can gain kinetic energy
by the 2D motion in the x-y plane than the 1D motion along the z-direction.
In fact, as shown in Fig.~3(d), there is a large density of
$d_{x^{2}-y^{2}}$ orbitals with a ferromagnetic-arrangement on the surface,
while $n_b(1)$ is relatively small.
Namely, electrons on the surface continue to occupy $d_{x^{2}-y^{2}}$
and escape only from  $d_{3z^{2}-r^{2}}$ orbitals,
leading to the overall increase of $N(1)$.

\section{Conclusion}

In this paper, we have evaluated $N(z)$,
the number of Mn$^{4+}$ ions per site in layer $z$,
by analyzing the orbital-degenerate double-exchange model with surfaces
for zigzag sheet AF and 3D FM states.
We have found that the surface-induced Friedel oscillations in $N(z)$ are enhanced, 
when the spinless $e_{\rm g}$-electron systems possess
Fermi-surface structures with improved nesting properties along the z-axis,
leading to an increase of Mn$^{4+}$ ions on the surface.
For cathodes and catalysts, retaining Mn$^{4+}$ ions on the surface
and Mn$^{3+}$ in the bulk is desirable.
Subsequently, we can conclude that 
the $t_{\rm 2g}$ spin structures possessing Fermi-surface structures
with good nesting properties are possibly suitable for high-efficiency cathodes in Li-ion batteries and catalysts. 
We believe that the present results provide a fundamental understanding for further application
of manganites to such systems.

\section*{Acknowledgment}

We thank K. Hattori, K. Kubo, and K. Ueda for discussions and comments.
The computation in this work was partly carried out using the facilities of the
Supercomputer Center of Institute for Solid State Physics, University of Tokyo.


\end{document}